\documentclass{ws-acs}
\usepackage{graphicx}
\usepackage{amsmath}
\usepackage{amssymb}
\usepackage[latin2]{inputenc}
\usepackage{color}
\def\slaninafigdir{.}
\begin{document}
\title{%
Complex temporal structure of activity in on-line electronic auctions 
}
\author{%
Franti\v{s}ek Slanina
}
\address{%
Institute of Physics,\\
 Academy of Sciences of the Czech Republic,\\
 Na~Slovance~2, CZ-18221~Praha,
Czech Republic
\\
slanina@fzu.cz
}

\maketitle
%

%
%
%
%
%
%
%
\begin{abstract}
We analyze empirical data from the internet auction site Aukro.cz. The
time series of activity   shows truncated fractal structure on scales 
from about 1 minute to about 1 day. The distribution of waiting times as 
well as the distribution of number of auctions within fixed interval
is a power law, with exponents $1.5$ and $3$, respectively. Possible
implications for the modeling of stock-market fluctuations are briefly
discussed.  
\end{abstract}


\section{Introduction}

Electronic markets emerged immediately after the rise of the
World-Wide Web. The initial 
enthusiasm fueled by the 
great expectations of the ``New economy'' was harshly put in its place
by the ``dot-com'' crash in the years 2000 to 2002, but in realistic view,
electronic commerce plays indeed ever increasing role within its own
segment of economy. There is nothing that would deserve the label
``New economy'', but there are plenty of new chances where the ``Old
good economy'' can get new blood. Among the various incarnations of
e-markets, we shall concentrate 
on a single segment, namely the on-line auctions. The best-known
example is the site Ebay.com, where virtually anybody can sell and buy
almost every legally accessible item. A good deal of other services throughout
the word imitate more or less successfully the Ebay.

Despite many studies devoted to it, some aspects of e-markets still
escape attention of the practitioners, perhaps because the questions are
considered too academic. However, we believe that asking fundamental,
``academic'' questions is beneficial for economic practice in the long
run. The agents' activity on electronic markets has been
studied for quite a long time now \cite{alt_kle_11}. 
Besides the analyses of overall structure of on-line auctions \cite{hou_blo_10},
much attention was devoted to the problem of winning strategy
\cite{rad_bar_ama_11,pig_ber_juu_gal_viv_11,yan_kah_06,wan_hu_09,sri_wan_10,pap_pie_10}
 and timing of the
placement of bids \cite{bor_boa_kad_06,shm_rus_jan_07,yan_kah_06,nam_sch_06}. The
study of network aspects of on-line auctions was initiated in the work
\cite{yan_jeo_kah_bar_03} and then it was investigated in depth 
\cite{yan_oh_kah_06,jan_yah_10}. One of the most important questions
asked was how the agents on the network cluster spontaneously
\cite{yan_oh_kah_06,rei_bor_07,pen_mul_08,srivastava_10,sla_kon_10}. 

The empirical studies of ``Old good economy'' concentrate mainly on
the amount and quality of fluctuations of prices
\cite{ma_sta_99,bou_pot_03}. Distribution of 
price changes and their correlations reveal well-known complex
patterns, the so-called ``stylized facts''.
 This
information is immediately used by speculators, so  that the knowledge
of them translates directly into money. 

On the other hand, fluctuations in electronic markets were much less
studied. In this work, we have in mind the fluctuations in the
intensity of trading. Although they may seem less practically
relevant, they are more academically appealing. Indeed, the complexity of
the noise in e. g. the stock market, is usually understood as a sign
of complex reaction of the system of large number of interacting
economic agents on a random external input, which may be a trivial
white noise. It is also supposed that feedback effects play
fundamental role and are behind the herding and other phenomena. In
short, an input, which may or may not be a Gaussian noise, is
processed in a complex way to give rise complex fluctuations seen in
stock markets.

On the contrary, the processing of random input signal is supposed
much weaker in electronic auctions. Essentially, most of the items
traded are independent. It can be also formulated as saying that the
ratio of number of commodities/number of agents is $\ll 1$ at stock
market, while $\gg 1$ at electronic auctions. 
Hence, the fluctuations are mostly due to the
dynamics of decisions of quasi-independent agents and therefore their properties
can be considered fairly close to the hypothetical input noise of the
stock market.

In this work we investigate the quantity and quality of fluctuations
which occur in trading at on-line auctions. The questions asked are
analogous to those well studied within the ``Old good economy'',
namely the fractal nature of the time series encoding the activity of
individual agents. Similarly to the ``stylized facts'' established in
stock markets, we try to grasp at least some of the ``stylized facts''
pertinent to electronic markets. Specifically, we shall analyze data
from the site aukro.cz, which is the Czech Republic division of the
multinational Allegro group.

\section{Description of the data set}

Most of the studies of on-line auctions were conducted on the
best-known Ebay, or on similar huge sites.
From practical point of view, less
prestigious sites  may be more attractive, because they often reveal
information which is hidden on Ebay and their smaller extent enables
collecting data sets which are ``full'' in the sense that they cover
nearly all activity on the site. We found that the site aukro.cz,
acting in the Czech Republic, has just the size we can effectively
download using a single dedicated computer.

We downloaded systematically all the information on the auctions that
were posted on the aukro.cz site, starting on 1 December 2009 and ending
18 November 2011. Due to occasional hardware and software problems the data are not
absolutely complete, but we estimate that the missing parts are not
larger than a few per cent of the existing data. In total, we have
information on $M=46,240,059$ auctions. Among them, there were
$M_\mathrm{rea}=11,473,486$ realized auctions, i. e those with at
least one bid. Each auction has a
unique seller and if  
realized, one or more bidders. We have information on unique ID numbers
of all sellers and bidders in all downloaded auctions. We identified
$N=1,083,276$ agents, among them
$N_\mathrm{sel}=460,867$ distinct sellers and
$N_\mathrm{bid}=1,004,703$ bidders. Of them,
$N_\mathrm{double}=382,294$ are double 
agents, i. e. are present
both in the set of sellers and in the set of bidders. 

Here we are interested in the properties of the time sequence of
activity of individual sellers. For each auction we know its ending
time. In the early stages of the project, we systematically recorded
the ended auctions. With this technology, the starting time and price
is lost. Only later we started recording the auction both at start and
end. Therefore, starting time sequences we have are shorter and we do not
analyze them here.

Thus, for each member of the set of sellers
 we construct the sequence of ending times
of all  auctions where she acted as a seller. (Note that the same
could be done also for bidders and more complicated structures can be
also studied, for example combined bidder ans seller sequences for
double agents.) For the seller $a$, $a=1,2...,N_\mathrm{sel}$, we
obtain the sequence of times $T_{i,a}$, 
$i=1,2,...,M_{a}$. In all the analysis, the unit of
measurement is one day.
We denoted $M_{a}$ the number of auctions posted by the
  seller $a$. We found that the maximum  is $\max_a M_a=150,916$, so that
   the time sequences can be fairly long. 
In the following we shall denote
$T_{\mathrm{min},a}=T_{1,a}$  the starting time of the sequence
corresponding to the seller $a$,  $T_{\mathrm{max},a}=T_{M_a,a}$ the
ending time and  $T_{\mathrm{span},a}=T_{\mathrm{max},a}-T_{\mathrm{min},a}$
the entire time span covered by the series.

\begin{figure}[t]
\includegraphics[scale=0.85]{%
\slaninafigdir/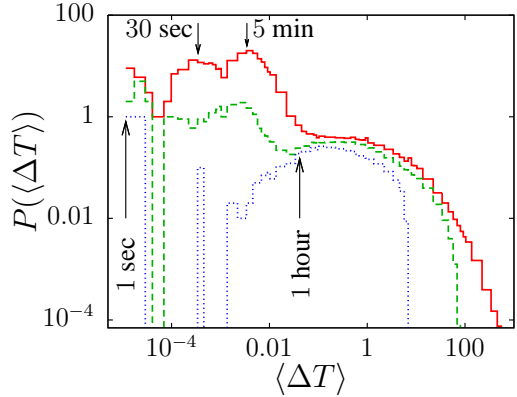}
\caption{Histogram of average distances between consecutive
  auctions of the same seller.
 Full line corresponds to all
  sellers, dashed line to sellers who initiated at least $10$ auctions,
  dotted line to sellers who initiated at least $100$ auctions. 
Time is given in days. The arrows
  indicate the positions of the features at $\langle\Delta
  T\rangle=1$~hour, $5$~minutes, $30$~seconds, and the minimum
  measured value $\langle\Delta
  T\rangle=1$~sec.}
\label{fig:distribution-aver-dist}
\end{figure}
\begin{figure}[t]
\includegraphics[scale=0.85]{%
\slaninafigdir/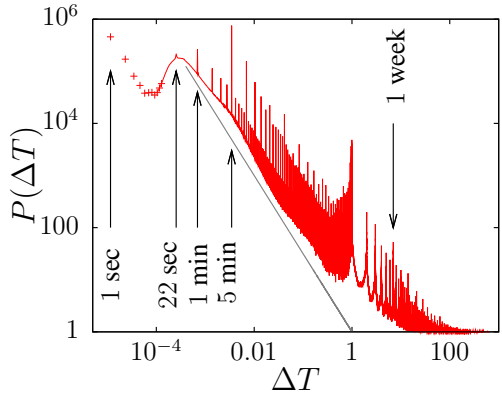}
\caption{Distribution of distances between consecutive
  auctions of the same seller. 
The data are summed for all time series in the
  data set. The unit of time is one day. The arrows indicate the
  positions of the  maxima at $\Delta T=1$~sec, $22$~secs, $1$~min, 
  $5$~minutes, and $1$~week.
  The straight line is the
  power-law dependence $\propto (\Delta T)^{-1.5}$. For $\Delta
  T>1$~day, the distribution is averaged over intervals $(\Delta
  T,\Delta T+1\;\mathrm{min})$, in order to see better the structure
  of peaks at multiples of days.}
\label{fig:distribution-distances}
\end{figure}

\section{Analysis of the time series}

When we analyze the time series of auctions, 
the first thing to look at is the average density in which the
auctions appear. Each seller acts with different mean frequency, so we
should ask first, what is the  distribution of the average
distances between auctions 
$\langle \Delta T\rangle_a=T_{\mathrm{span},a}/(M_a-1)$. 
The results are shown in
Fig. \ref{fig:distribution-aver-dist}. We can see that most sellers
act at a scale up to  $1$~day, with a broad tail extending up to
several hundred days, which is about the total extent of our data. 
So,
the tail is to a large extent determined by the finite observation
time. The
maximum of the distribution lies around $\langle \Delta T\rangle\simeq
5$~minutes and it is not simple, but fairly broad an 
structured, signaling
several types of sellers. So, the set of sellers is very
heterogeneous.

However, if we average the distances between auctions over all sellers, we get
quite a long time, about $30$ days. This means that the peaks at short
distances are somehow not typical. To check this, we looked at what
types of sellers constitute these short-distance peaks. We found that
mostly they are occasional sellers, who tried to sell a chunk of items
within very short time and then they never acted again. Most of them
sold at most 10 items. Among about $8000$ sellers characterized by
average distance between auctions at most $10$~min, we identified just
$13$ intense traders, i. e. those with at least $1000$ auctions. They
specialized in several typical branches, like used books (at least $4$
of these $13$ sellers), cosmetics, collectibles (old postcards,
fossils, etc.), bijou, and the like. To filter out the occasional
sellers, we plot in Fig. \ref{fig:distribution-aver-dist} also two
other histograms,
 restricted to sellers with at least $10$ and at least $100$
 auctions. We can clearly see that the peaks at very small average time
 distances substantially diminish in the former and vanish in the
 latter histogram. But still the set of sellers with at least $100$
 auctions cover scales of $\langle T\rangle$ starting at $5$ minutes
 and ending at tens of days. This is a fairly wide range of scales.

When we look at the time series  (examples
will be shown soon), we observe complex patterns. The first step in
understanding the patterns is to see the distribution of the distances
between auctions 
\begin{equation}
P(\Delta T)=\sum_a\sum_{i=1}^{M_a-1}\delta\Big(T_{a,i+1}-T_{a,i}-\Delta
T\Big)\;.
\end{equation}
(Recall that the time series belongs to a single seller, so the
distance is the time which separates the consecutive auctions of the
same seller.)
The distribution is shown in
Fig. \ref{fig:distribution-distances}. We can see large number of
sharp peaks at special values of the time distances. However, all of
them are artifacts in the sense that they correspond to multiples of
natural units of time in which the sellers organize their
activity, namely minutes, multiples of $5$~minutes, and especially
multiples of days. The non-trivial feature is rather the background decay, over
which the peaks are superposed. The data in Fig.
\ref{fig:distribution-distances} suggest that within the time scale of
minutes to one day, the background decay is a power law $P(\Delta T)\sim
(\Delta T)^{-\gamma}$ with $\gamma\simeq 1.5$.

\begin{figure}[t]
\includegraphics[scale=0.85]{%
\slaninafigdir/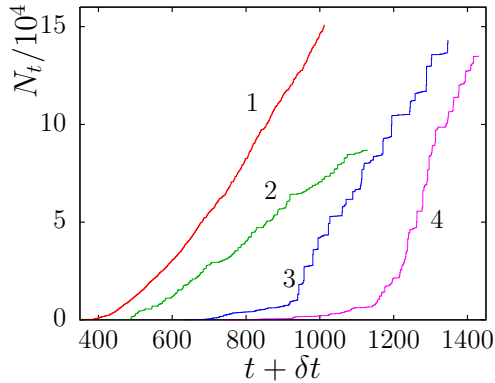}
\caption{Four typical examples of the time series of activity of
  sellers on the Aukro site. The numbers $1$ to $4$ refer to
  corresponding data in Figs. \ref{fig:timeseries-examples-detail} and
  \ref{fig:fractal-examples}. Time is given in days, day $0$ is 1
  January 2009. For better visibility, the time was shifted by the
  interval $\delta t=0$ for curve $1$, $\delta t=100$ for curve $2$,
  $\delta t=300$ for curve $3$, and $\delta t=400$ for curve $4$. }
\label{fig:timeseries-examples}
\end{figure}
\begin{figure}[t]
\includegraphics[scale=0.85]{%
\slaninafigdir/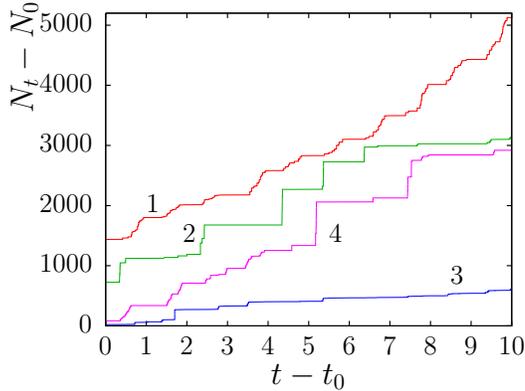}
\caption{Details of the timeseries shown in
  Fig. \ref{fig:timeseries-examples}, covering $10$ days. The starting
day was $t_0=750$ for curves $1$, $2$, and $4$, and $t_0=970$ for
curve $3$. For better visibility, the curves were shifted downwards by
$N_0=65000$, $47500$, $116200$, and $7600$ for curves $1$, $2$, $3$,
and $4$, respectively. Time is given in days.}
\label{fig:timeseries-examples-detail}
\end{figure}

Let us turn to the analysis of the individual time series now. 
The overall impression on the time series can be glimpsed from the
cumulative number of auctions for one seller up to time $t$, i. e.
\begin{equation}
N_{t,a}=\sum_{i=1}^{M_{a}}\theta(t-T_{i,a})
\end{equation}
where $\theta(x)=1$ for $x>0$ and $\theta(x)=0$ elsewhere. We show in
Fig. \ref{fig:timeseries-examples} four examples of such time
series. We can see that they  look rather different. Some of them
appear fairly smooth, other exhibit marked steps. Details of the same
time series, covering interval of just 10 days is shown in
Fig. \ref{fig:timeseries-examples-detail}. We can again observe steps at  
this smaller scale, indicating that a structure close to a fractal is
present at least in some of the series. Among several methods of
checking the fractality of the series we chose the standard
box-counting method, i. e.
covering the time span of the series by intervals of variable length. 

Let us take one time series, for seller $a$. 
We fix for a moment a time interval $\Delta t $ and divide the time span 
$T_{\mathrm{span},a}=T_{\mathrm{max},a}-T_{\mathrm{min},a}$
 into non-overlapping intervals of
length  $\Delta t$. Then, we count the number $N_{\mathrm{occup},a}(\Delta
t)$ of
such intervals containing at least one auction. Of course, we do that
only for $\Delta t < T_{\mathrm{span},a}$.

\begin{figure}[t]
\includegraphics[scale=0.85]{%
\slaninafigdir/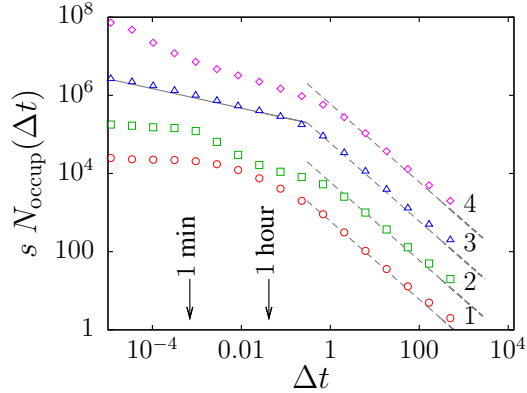}
\caption{Number of intervals of length $\Delta t$, containing at least
  one auction, for the timeseries shown in
  Fig. \ref{fig:timeseries-examples}. Time is given in days, arrows
  showing the scale of one minute and one hour. For better visibility,
the data are shifted upwards by a factor $s=1$, $10$, $100$, and
$1000$, for curves $1$, $2$, $3$,
and $4$, respectively. The straight lines are power dependencies
$\propto(\Delta t)^{-0.25}$ (solid line), and $\propto(\Delta t)^{-1}$ (dashed lines).}
\label{fig:fractal-examples}
\end{figure}

We can see in Fig.
\ref{fig:fractal-examples}
the dependence $N_{\mathrm{occup},a}(\Delta t)$ for the same four examples
of time series as shown in  Fig \ref{fig:timeseries-examples}. (We
drop the index $a$ in the axis label, as it bears no information
there.)
 Several
regimes can be identified. At times longer than about one day, all the
four time series exhibit the behavior 
$N_{\mathrm{occup}}(\Delta t)\propto (\Delta t)^{-1}$, typical for
fractal dimension $\kappa=1$. At shorter time scales, the behavior is
much different and non-universal. For example, the time series No. 2
exhibits two shoulders, at $1$ day and at $1$ minute. There is a clear
plateau below $1$ minute, showing that at such small scales the
fractal dimension is zero. Less clear ``plateau'' can be seen between
$1$ minute and $1$ day, which may perhaps be better characterized as a
depression at scale about $1$ hour. This suggests that two typical scales are
there. The auctions are put no closer than about $1$ minute, but in
bunches extending no more than about one hour. On a scale larger than
about one day, the time series looks  uniform.
This is in marked
contrast with time series No. 3 and No 4. At scales shorter than $1$
day, the auctions seem to form a set close to a fractal, more clear for No. 3,
with fractal dimension $\kappa\simeq 0.25$, and less clear for No. 4,
with slightly larger fractal dimension. The time series No. 1 is yet
another type. Again, there are no structures in the time series below
the scale of $1$ minute and beyond $1$ day the time series is
uniform. However, at the intermediate scales, there is a broad
crossover, with neither a  clear separation of time scales nor any sign of
fractality. 

Despite the diversity of the characters of the time series we looked
also on their average properties. However, we have to be careful in making
the average, for several reasons. First, the time span of the time
series varies and so varies the maximum interval $\Delta t$ which
enters into the time span. If we averaged directly the number of
occupied intervals, we would introduce a systematic error. Therefore,
we average the normalized quantity
\begin{equation}
\overline{N_\mathrm{occup,norm}(\Delta t)}
=\Big(\sum_a\theta(\Delta t-T_{\mathrm{span},a})\Big)^{-1}
\sum_a\theta(\Delta t-T_{\mathrm{span},a})
\frac{N_{\mathrm{occup},a}(\Delta t)}{T_{\mathrm{span},a}}\;.
\label{eq:average-noccup}
\end{equation}
Second, we found that time sequences with too short time span or
containing too few auctions introduce spurious artifacts. Therefore,
we limit the sum over $a$ in (\ref{eq:average-noccup}) to such series,
for which time span extends more than one week,
i. e. $T_{\mathrm{span},a} > 7$, and which contain at least $300$
auctions, i. e. $M_{a}\ge 300$.

\begin{figure}[t]
\includegraphics[scale=0.85]{%
\slaninafigdir/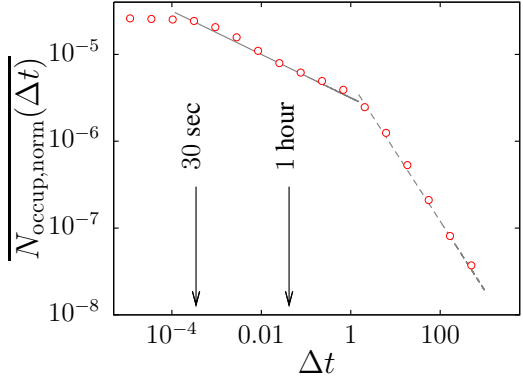}
\caption{Number of intervals of length $\Delta t$, containing at least
  one auction, normalized to the time span, and 
averaged over all timeseries in the data set longer than one week and
  containing at least $300$ auctions. Time is given in days, arrows
  showing the scale of one minute and one hour. The straight lines are
  power dependencies 
$\propto(\Delta t)^{-0.25}$ (solid line), and $\propto(\Delta
  t)^{-0.8}$ (dashed line).} 
\label{fig:fractal-all}
\end{figure}

The result is shown in Fig. \ref{fig:fractal-all}. After averaging,
three regimes are clearly visible. For the scales below 
$\Delta t_1\simeq 30$~seconds, the fractal dimension is zero on average, indicating
that typical time sequences contain no structure at times shorter than
$\Delta t_1$. Between $\Delta t_1 $ and the longer scale $\Delta
t_2\simeq 1$~day, the power-law decay suggests fractal structure with
dimension $\kappa\simeq 0.25$. On longer scales than $\Delta t_2 $,
the decay follows a power law with exponent $\kappa\simeq 0.8$,
somewhat lower than $1$, showing some non-trivial structure also at
longer scales. This result deviates from the data shown in
Fig. \ref{fig:fractal-examples}, where all four samples are fitted to
$\kappa\simeq 1$ at long time scales, but the deviation is rather
small.  This indicates that some time 
series are structured also at larger times, but it is not common.

\begin{figure}[t]
\includegraphics[scale=0.85]{%
\slaninafigdir/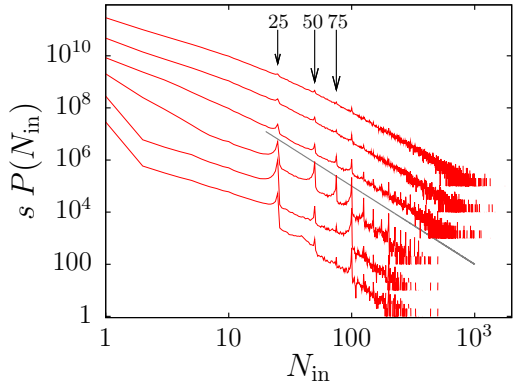}
\caption{Distribution of number of auctions $N_\mathrm{in}$ contained
  within a time interval $\Delta t$, summed for all time series in the
  data set. The six curves correspond to intervals (from
  the bottom to the top) $\Delta t = 1$~sec, $10$~sec, $100$~sec,
  $10^3$~sec, $10^4$~sec, and $10^5$~sec. For better
  visibility, the data are shifted upwards by the factors (from
  the bottom to the top) $s=1$, $10$, $100$, $10^3$, $10^4$, and $10^5$.
The arrows indicate the
  positions of $N_\mathrm{in}=25$, $50$, and $75$, to stress that
  these are just the positions of the first peaks. The straight line is the
  power-law dependence $\propto N_\mathrm{in}^{-3}$.}
\label{fig:nu-per-deltat}
\end{figure}
\begin{figure}[t]
\includegraphics[scale=0.85]{%
\slaninafigdir/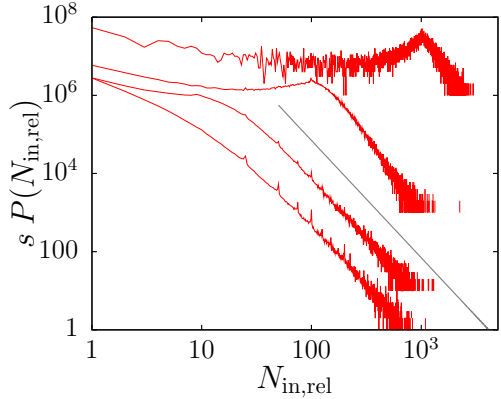}
\caption{Distribution of number of auctions $N_\mathrm{in,rel}$ contained
  within a time interval $\Delta t=k\langle\Delta T\rangle$, where
  $\langle\Delta T\rangle$ is the average distance between consecutive
  auctions in the same time series. 
  The data are summed for all time series in the
  data set. The four curves correspond to multiples (from
  the bottom to the top) $k=1$, $10$, $100$ and $1000$. For better
  visibility, the data are shifted upwards by the factors (from
  the bottom to the top) $s=1$, $10$, $10^3$, and $10^6$.
The straight line is the
  power-law dependence $\propto N_\mathrm{in,rel}^{-3}$.}
\label{fig:nu-per-deltat-relative}
\end{figure}

The box counting method used above does not reveal all the complexity. A
complementary method consists in plotting distribution of number of
auctions present within an interval of fixed length. We divide the time
span into intervals  $(t_{l,a},t_{l+1,a})$,
$t_{l,a}=T_{\mathrm{min},a}+l\Delta t$ 
for $0\le l< T_{\mathrm{span},a}/\Delta t$ and count  
\begin{equation}
P(N_\mathrm{in})=
\sum_a\sum_k\delta\Bigg(\sum_{i=1}^{M_a}
\theta(T_{i,a}-t_{l,a})\theta(t_{l+1,a}-T_{i,a})
-N_\mathrm{in}\Bigg) \;.
\label{eq:def-nin-statistics}
\end{equation}
The
results are shown in Fig. \ref{fig:nu-per-deltat} for intervals of
lengths $\Delta t=10^n$ seconds, $n=0,1,2,3,4,5$. We can see that all
these interval lengths lead to the same general form of the
distribution, with a tail following a power law $P(N_\mathrm{in})\sim
N_\mathrm{in}^{-\tau}$ with exponent $\tau\simeq 3$. However, on top
of this background power law,
 there are significant structures of peaks. The peaks are highest
 at the shortest interval $\Delta t=1$~sec
and they gradually vanish when $\Delta t$ increases. The most interesting
fact is that the positions of the
peaks are at integer multiples of $25$, the highest ones being at
$N_\mathrm{in}=25$ and $N_\mathrm{in}=100$. This is clearly due to
large-scale sellers which place large chunks of auctions
at one moment (within one second). More detailed view shows that
for $\Delta t=1$~s the
peaks at $50$ and $75$ are smaller than those at $25$ and $100$, while
for $\Delta t=100$~s these four peaks are of comparable weight, if
measured with respect to the background power law, and also the peaks
at multiples of $100$ are more pronounced than in the distribution
corresponding to $\Delta t=1$~s. This means that
there are sellers who put several chunks of sizes $25$, or $100$ not
exactly at the same second, but still not farther than about one minute.

In the distribution $P(N_\mathrm{in})$, as shown in Fig.
\ref{fig:nu-per-deltat}, we mix sellers who act with different
intensity.
 We have already seen the distribution of the average
distances between auctions in
Fig. \ref{fig:distribution-aver-dist}. 
In order to take into account the observed heterogeneity in average
distance between auctions, we modify the statistics  defined by
(\ref{eq:def-nin-statistics})
 so that the time span of seller $a$ is divided into intervals of length $\Delta
 t=k\langle T\rangle_a$,  i. e. $t_{l,a}=T_{\mathrm{min},a}+lk\langle
 T\rangle_a$. The results are shown in
 Fig. \ref{fig:nu-per-deltat-relative}.  Again, we can observe the
 power-law tail with exponent close to $\tau\simeq 3$, but a peak
 develops at the value $N_\mathrm{in,rel}=k$, as should be expected,
 because $k$ is exactly the average value for the number of auctions
 within the interval  $\Delta  t=k\langle T\rangle_a$, for every
 $a$. The exponent of the power-law tail seems to decrease when $k$
 increases, but the power-law dependence is still visible. 
We can conclude that the power-law tail with exponent $\tau\simeq 3$
is due to intrinsic structure of the time series, rather than
heterogeneity of the average time scales. Therefore, the noise induced
by the activity of the individual agents is itself complex.

\section{Conclusions}

We performed analysis of complex structure of activity in on-line
auctions. For each individual seller, we looked at the time series of
the auctions she initiated. The basic, though naturally expected,
 fact is that the set of sellers
is very heterogeneous. The average distance between auctions of the
same seller varies from seconds to hundreds of days. 

The distribution of distances between subsequent auctions of the same
seller exhibits fundamental power-law decay on the scale from
$1$~minute to $1$~day, with exponent $\simeq 1.5$. On top of this
power-law, there are peaks at natural units of time, like one minute
or multiples of $5$~minutes. 

We tested the signs of fractality in the time series of
auctions. Although each series is unique, the common features can be
identified. We must note in advance that the word ``fractality'' is
somewhat abused here, as we always talk on fractal properties within
a limited range of scales. By the box counting method, we found that
beyond the scale of one day, the series are uniform, i. e. their
fractal dimension is $1$. On the other end of the scales, for times
shorter than about $30$~seconds the series appears as a set of
isolated points, since the measured fractal dimension is $0$. In the
intermediate range, from $30$~sec to $1$~day, self-similar structure
is observed, with fractal dimension $\simeq 0.25$. Note, however, that
these measurements describe averaged properties of the time series,
but individual time series may deviate from this mean
behavior. Indeed, we observed individual time series with clear
fractal shape, and others, in which just two typical scales were
observed, without signs of fractality.

From the other side, we found the statistics of number  of auctions
within an interval of fixed length. This length was either
equal for all sellers, or a fixed multiple of average distance between
auctions, i. e. different for each seller. In both cases we observed a
power-law tail with exponent $\simeq 3$, just the value at the border
between L\'evy stable and Gaussian distributions. So, the tail is not
due to heterogeneity of the set of sellers, but due to the dynamics of
each seller separately.

In all the analyses we observed complex patterns of noise in the
placement of auctions by individual sellers. There is power-law
statistics in the waiting times between actions of the agents, there
is power law in the statistics of cumulative activity of an agent over
fixed time interval. We may speculate on the
consequences of these findings for possible modeling of the behavior
of agents in, for example, stock markets. If the agents issued orders
to buy or sell shares, rather than placed auctions, the cumulative
effect of the power-law distributed numbers of orders would be a
power law slowly converging to a Gaussian, which is just the general
view of how the statistic of stock market price changes looks like.

Finally, let us mention that similar study of buyer
behavior was done. The results seem to be fairly similar, which is
probably due to the fact that content and timing of the auctions is
entirely in hands of the sellers and the buyers just follow what they
could. More subtle differences between the sellers' and buyers' side
and their possible synchronization will be investigated in a future study.

\section*{Acknowledgments}
I wish to thank Z. Konop\'asek for numerous fruitful discussions.
This work was carried out within the project AV0Z10100520 of the Academy 
of Sciences of the Czech republic and was  
supported by the M\v{S}MT of the Czech Republic, grant no. 
OC09078.

\end{document}